\documentclass[aip,prl,twocolumn]{revtex4}
\usepackage{graphics,graphicx,color}
\begin{document}

\title[]{Damping and softening of transverse acoustic phonons in colossal magnetoresistive La$_{0.7}$Ca$_{0.3}$MnO$_3$ and La$_{0.7}$Sr$_{0.3}$MnO$_3$}
\author{Joel~S.~Helton$^{1,2,\ast}$, Yang~Zhao$^{2,3}$, Dmitry~A.~Shulyatev$^{4}$, and Jeffrey~W.~Lynn$^{2,\dag}$}

\address{$^{1}$Department of Physics, United States Naval Academy, Annapolis, MD 21402, USA}
\address{$^{2}$NIST Center for Neutron Research, National Institute of Standards and Technology, Gaithersburg, MD 20899, USA}
\address{$^{3}$Department of Materials Science and Engineering, University of Maryland, College Park, MD 20742, USA}
\address{$^{4}$National University of Science and Technology ``MISiS", Moscow 119991, Russia}

\begin{abstract}
Neutron spectroscopy is used to probe transverse acoustic phonons near the $(2, 2, 0)$ Bragg position in colossal magnetoresistive La$_{0.7}$Ca$_{0.3}$MnO$_3$ and La$_{0.7}$Sr$_{0.3}$MnO$_3$. Upon warming to temperatures near $T_c=257$~K the phonon peaks in La$_{0.7}$Ca$_{0.3}$MnO$_3$ soften and damp significantly with the phonon half width at half maximum approaching 2.5~meV for phonons at a reduced wave vector of $\vec{q}=(0.2, 0.2, 0)$. Concurrently a quasielastic component develops that dominates the spectrum near the polaron position at high temperatures. This quasielastic scattering is $\approx5$ times more intense near $T_c$ than in La$_{0.7}$Sr$_{0.3}$MnO$_3$ despite comparable structural distortions in the two. The damping becomes more significant near the polaron position with a temperature dependence similar to that of polaron structural distortions. An applied magnetic field of 9.5~T only partially reverses the damping and quasielastic component, despite smaller fields being sufficient to drive the colossal magnetoresistive effect. The phonon energy, on the other hand, is unaffected by field. The damping in La$_{0.7}$Sr$_{0.3}$MnO$_3$ near $T_c$ at a reduced wave vector of $\vec{q}=(0.25, 0.25, 0)$ is significantly smaller but displays a similar trend with an applied magnetic field.

\end{abstract}
\maketitle

\section{I. Introduction}

Perovskite manganites of the form $R_{1-x}A_x$MnO$_3$, where $R$ is a trivalent rare-earth cation and $A$ is a divalent alkaline-earth cation, are important systems for exploring the interplay of magnetic, electrical, lattice, and orbital degrees of freedom \cite{RamirezJPCM1997,RaveauRao1998}. La$_{1-x}$Ca$_x$MnO$_3$ has attracted particular attention \cite{SchifferPRL1995,EdwardsAP2002}. At half doping La$_{0.5}$Ca$_{0.5}$MnO$_3$ displays an antiferromagnetic ground state with CE-type charge and orbital order \cite{WollanPRW1955,GoodenoughPRW1955} that results in superlattice peaks with a $\vec{k}=(0.25, 0.25, 0)$ propagation vector. At lower doping the ground state is a ferromagnetic metal, and colossal magnetoresistance  (CMR) is observed at the combined ferromagnetic and metal-insulator phase transition. Double exchange alone is insufficient to explain the magnitude of the CMR effect \cite{MillisPRL1995}, with a role played by polarons in the insulating phase where Jahn-Teller interactions cause local CE-type distortions that trap charge carriers \cite{MillisPRL1996,MillisPRB1996}.

La$_{0.7}$Ca$_{0.3}$MnO$_3$ displays a fairly significant CMR effect and evidence for both static \cite{AdamsPRL2000,DaiPRL2000} and dynamic~\cite{LynnPRB2007} polarons. Scattering near the $\vec{q}=(0.25, 0.25, 0)$ polaron position arises from correlated polarons and peaks near $T_c$. Diffuse scattering from uncorrelated polarons rises rapidly in intensity as the temperature approaches $T_c$ but remains constant at higher temperatures, suggesting that the number of polarons remains relatively constant above $T_c$ even while their correlations decrease at higher temperature. A polaron glass phase persists to 400~K. Ridges of quasielastic magnetic scattering \cite{HeltonPRB2012,HeltonPRB2014} associated with the magnetic part of diffuse polarons are only partially suppressed by a 10-T magnetic field, indicating that very large fields are needed to change the number of polarons near $T_c$. An optical Jahn-Teller phonon mode displays anomalous and increasing damping with increasing temperature and collapses above $T_c$~\cite{ZhangPRL2001}. La$_{0.7}$Sr$_{0.3}$MnO$_3$, with a higher transition temperature, displays a less pronounced CMR effect \cite{UrushibaPRB1995}, though there is evidence of significant structural distortions at $T_c$ \cite{LoucaPRB1997,WeberPRB2013} and a polaronic metal phase \cite{MaschekPRB2016}. Unlike La$_{0.7}$Ca$_{0.3}$MnO$_3$, this compound displays only dynamic polarons demonstrating that the strength of the colossal magnetoresistive effect is correlated with the polaron lifetime \cite{MaschekPRB2016,MaschekPRB2018}.

\section{II. Experiment and Results}

La$_{0.7}$Ca$_{0.3}$MnO$_3$ (LCMO) and La$_{0.7}$Sr$_{0.3}$MnO$_3$ (LSMO) have an orthorhombic perovskite crystal structure; however, the orthorhombic distortions are small relative to the resolution of neutron spectroscopy measurements and the crystallographic domains are equally populated so we index the samples with a pseudocubic notation where $a=b=c\approx3.9$~{\AA}. Polaron scattering has been observed at a reduced wave vector of $\vec{q}=(\pm0.25, \pm0.25, 0)$ in materials such as La$_{0.7}$Ca$_{0.3}$MnO$_3$ \cite{AdamsPRL2000,DaiPRL2000} and the bilayer manganite La$_{2-2x}$Sr$_{1+2x}$Mn$_2$O$_7$ \cite{ArgyriouPRL2002}, consistent with the superlattice propagation vector of CE-type charge and orbital order \cite{GoodenoughPRW1955}. The polaron scattering structure factor is particularly intense for reduced wave vectors transverse to the (2,~2,~0) Bragg position: $\vec{Q}=(2.25,~1.75,~0)$ and equivalent. Both longitudinal and transverse phonons with $\vec{q}=(\pm0.25, \pm0.25, 0)$ feature oxygen displacements that partially match the distortions of CE-type order~\cite{WeberNM2009}. The structure factor for transverse acoustic phonons is also quite large around the (2,~2,~0) Bragg peak. Therefore we have concentrated this study on the spectroscopy of transverse acoustic phonons at positions equivalent to $\vec{Q}=(2+\xi,~2-\xi,~0)$ with $\xi$ between 0.10 and 0.25. Because the sample contains multiple crystallographic domains, at a reduced wave vector of (0.25,~0.25,~0) in the pseudocubic notation, only a portion of the sample will be oriented at the actual polaron position of $\vec{q}=(0.5,~0,~0)$ in the orthorhombic notation. A previous work found that the damping of transverse acoustic phonons in LSMO \cite{MaschekPRB2016} increased significantly upon approaching the nominal polaron position with $\xi$=0.25 but was maximized at a larger value close to $\xi=0.3$, possibly related to the averaging over multiple domains.  

Neutron spectroscopy measurements utilized the BT-7 thermal triple-axis spectrometer \cite{BT7} at the NIST Center for Neutron Research with Open-50$^\prime$-50$^\prime$-120$^\prime$ collimators, a fixed final energy of $14.7$~meV, and pyrolytic graphite filters after the sample. Constant-$Q$ scans were measured for transverse acoustic phonons at $\vec{Q}=(2-\xi, -2-\xi,0)$ in LSMO for $\xi$=0.10, 0.15, 0.20, 0.25, and 0.30 at temperatures of $T$=250~K, 275~K, 300~K, 325~K, 350~K, and 370~K. This LSMO sample had a mass of 3~g and $T_c=351$~K \cite{VasiliuDolocJAP1998}. Constant-$Q$ scans were measured for transverse acoustic phonons at $\vec{Q}=(2+\xi, -2+\xi,0)$ in LCMO for $\xi$=0.10, 0.15, 0.20, and 0.25 at temperatures of $T$=100~K, 150~K, 200~K, 250~K, and 300~K. This LCMO sample has a mass of 1.5~g and $T_c=257$~K (higher than most other single-crystal samples of LCMO) \cite{ShulyatevJCG2002}. Further constant-$Q$ scans were measured with the samples placed in a vertical field superconducting magnet. For LCMO a 10-T magnet was used and scans were measured at $\vec{Q}=(1.85, -2.15, 0)$ and $\vec{Q}=(1.80, -2.20, 0)$ at 265~K for magnetic fields of 0~T, 3~T, 6~T, and 9.5~T as well as at 140~K in zero field.  For LSMO a 7-T magnet was used and scans were measured at $\vec{Q}=(1.75, -2.25, 0)$ at 345~K for magnetic fields of 0.5~T, 3~T, and 6~T as well as at 250~K in zero field. A sloped background has been fit and then removed from data taken inside the magnet. Error bars and uncertainties throughout this paper are statistical in nature and represent one standard deviation.

\begin{figure}
\centering
\includegraphics[width=8.3cm]{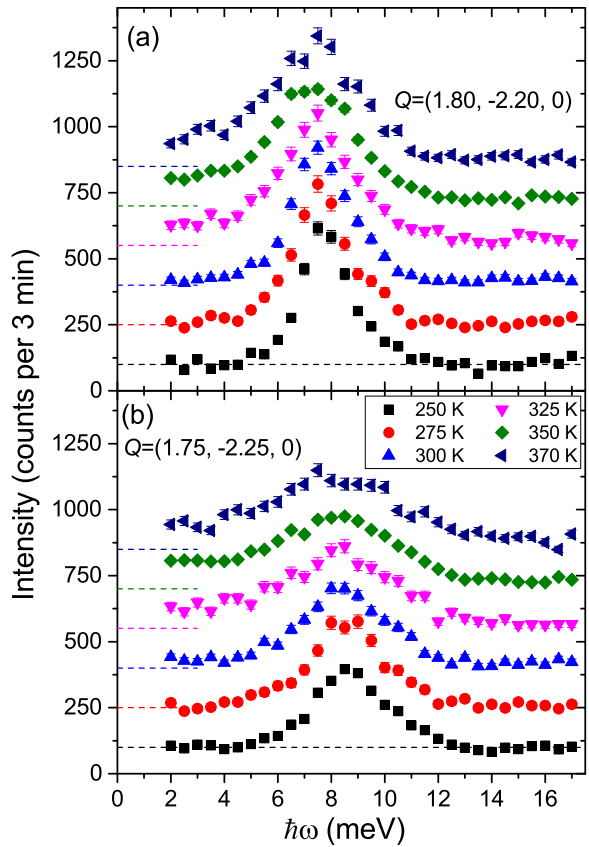} \vspace{-5mm}
\caption{Constant-$Q$ scans of transverse acoustic phonons in La$_{0.7}$Sr$_{0.3}$MnO$_3$ ($T_c=351$~K) at a series of temperatures. (a) $\xi=0.20$ scan at $\vec{Q}=(1.80, -2.20, 0).$ (b) $\xi=0.25$ scan at $\vec{Q}=(1.75, -2.25, 0).$} \vspace{-2mm}
\label{LSMO_0p20_0p25}
\end{figure}

Figure~\ref{LSMO_0p20_0p25} displays constant-$Q$ scans of the transverse acoustic phonons in LSMO at $\xi=0.20$ and $\xi=0.25$ (the polaron position). For both positions a qualitative change is observed upon heating above 300 K ($\approx0.85T_c$). The phonons become broadened, with extra scattering intensity observed on both the high-energy and low-energy sides of the peak. The extra scattering on the low-energy side of the peak is particularly pronounced and extends to the lowest energy transfers measured. Maschek~\emph{et~al.}~\cite{MaschekPRB2016} have reported this behavior in LSMO modelled as a phonon peak that softens and damps with increasing temperature alongside a quasielastic component arising from dynamic polarons.

\begin{figure*}
\centering
\includegraphics[width=18.5cm]{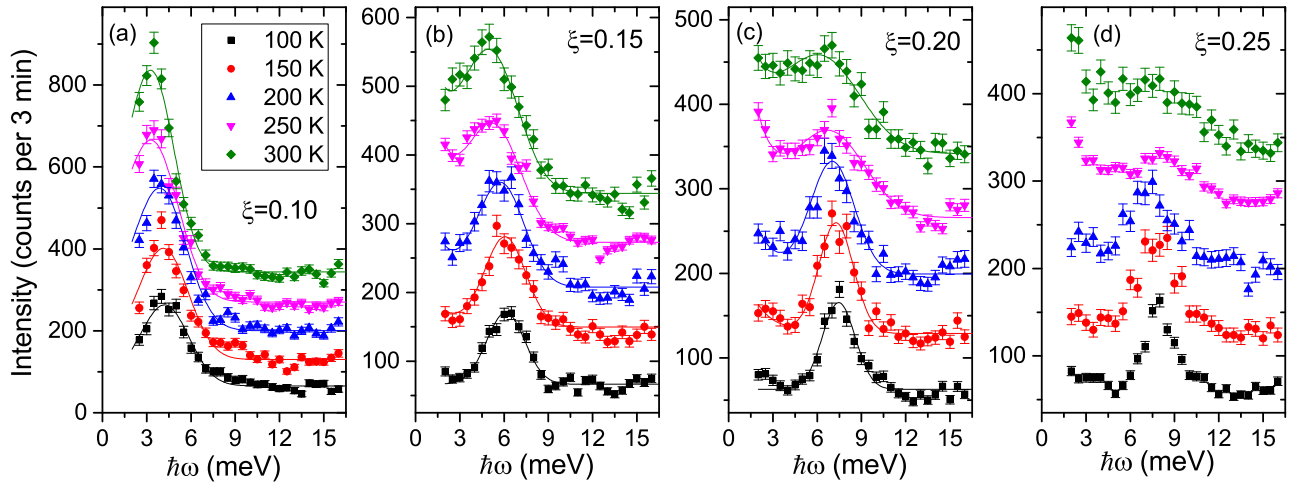} \vspace{-8mm}
\caption{Constant-$Q$ scans of transverse acoustic phonons in La$_{0.7}$Ca$_{0.3}$MnO$_3$ ($T_c=257$~K) at a series of temperatures. (a) $\xi=0.10$ scan at $\vec{Q}=(2.10, -1.90, 0).$ (b) $\xi=0.15$ scan at $\vec{Q}=(2.15, -1.85, 0)$. (c) $\xi=0.20$ scan at $\vec{Q}=(2.20, -1.80, 0).$ (d) $\xi=0.25$ scan at $\vec{Q}=(2.25, -1.75, 0).$ Successive data sets are offset for clarity. The solid lines in the first three panels are fits as described in the text. } \vspace{-2mm}
\label{LCMO_Data}
\end{figure*}

Figure~\ref{LCMO_Data} displays constant-$Q$ scans in LCMO at temperatures between 100~K and 300~K for phonons with $\xi=$0.10 to 0.25 (the polaron position). Qualitatively similar behavior is observed with phonon peaks that soften and damp upon warming through 200~K ($\approx0.78T_c$). However, the quasielastic component is significantly more intense compared to LSMO. For temperatures at 250~K and above the quasielastic component is sufficiently intense that it merges with the broadened phonon peak to yield intensity across a wide range of energy transfers on the low-energy side of the peak that are comparable in intensity to the peak position.

\begin{figure}
\centering
\includegraphics[width=8.3cm]{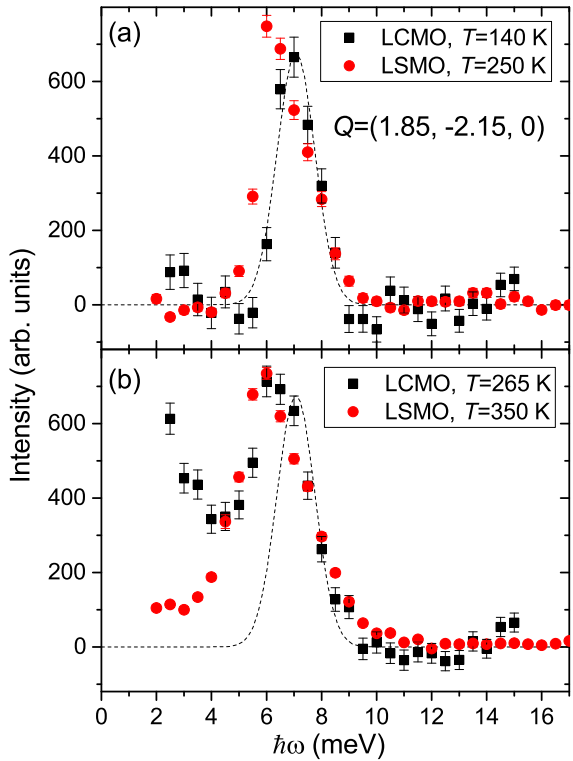} \vspace{-5mm}
\caption{Constant-$Q$ scans for both LCMO and LSMO at $\vec{Q}=(1.85, -2.15, 0)$. (a) Scans performed at temperatures well below $T_c$. The dashed line is a Gaussian fit to the LCMO data. (b) Scans performed close to $T_c$. The dashed line is the same function as in the top panel.} \vspace{-2mm}
\label{Comparison}
\end{figure}

Figure~\ref{Comparison} allows for a direct comparison of phonon damping and softening as well as the quasielastic scattering component in LCMO and LSMO. Constant-$Q$ scans are shown for both materials at $\vec{Q}=(1.85, -2.15, 0)$; this $\xi=0.15$ position is farther from the polaron position than some other data but allows for easier discernment between the phonon and quasielastic contributions. At temperatures well below $T_c$, shown in Fig.~\ref{Comparison}(a), the phonon is comparable in the two materials. The atomic mass difference between La and Sr is significantly less than the difference between La and Ca leading to a smaller average and greater variance of the $A$-site atomic mass in LCMO. Differing levels of $A$-site chemical disorder \cite{ShibataPRL2002} are also possible. Despite these differences the phonon spectra at $\xi=0.15$ are quite similar between the two compounds. The phonon peak is slightly softer and broader in LSMO, likely reflecting the higher temperature. Close to $T_c$ the phonon has softened and broadened to a comparable extent in both materials, as the phonon peaks shown in Fig.~\ref{Comparison}(b) still largely overlap. However the quasielastic intensity is far more pronounced in LCMO. In the LSMO data the quasielastic contribution is visible only as a plateau in intensity below 3~meV while in LCMO it presents as a peak with intensity comparable to the phonon. Fitting both data sets to a quasielastic component plus a phonon contribution we find that the quasielastic dynamic susceptibility in LCMO at 265~K is a factor of $5.3\pm1.2$ more intense than in LSMO at 350~K.

\begin{figure}
\centering
\includegraphics[width=8.3cm]{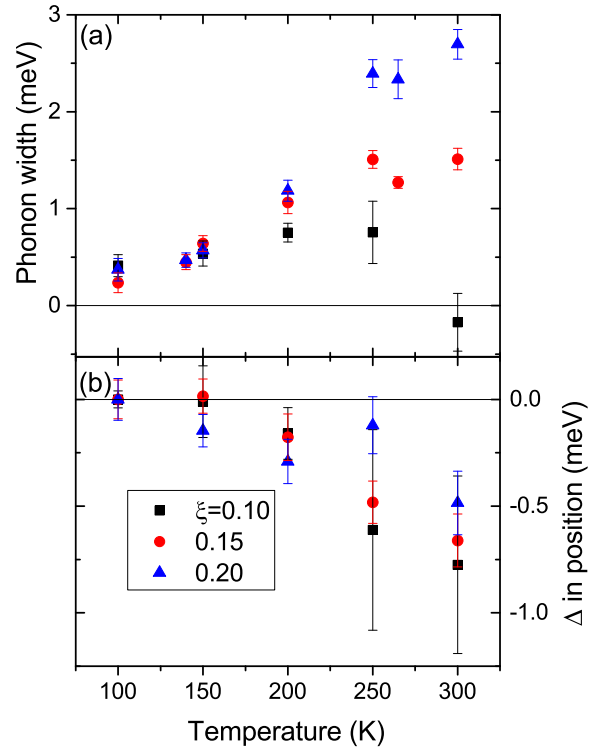} \vspace{-5mm}
\caption{Fit parameters for La$_{0.7}$Ca$_{0.3}$MnO$_3$ in zero field as a function of temperature. (a) Phonon width (HWHM) after deconvolution with the instrumental resolution. (b) Change in the phonon peak position from $T=100$~K. } \vspace{-2mm}
\label{LCMO_fits}
\end{figure}

The constant-$Q$ scans in LCMO were fit to the sum of a quasielastic contribution and a phonon peak. Fits were performed with the phonon modeled as both a Gaussian and a damped harmonic oscillator. Fit values for the two cases were similar; values reported here used a Gaussian line shape as it provided a slightly better fit to the data. For each $\vec{Q}$-position the lowest temperature data ($T=100$~K for zero-field scans, $T$=140~K for scans measured in the magnet) were fit first with the quasielastic peak set at zero intensity. This fit was used to fix the phonon dynamic susceptibility, with the temperature dependence of the integrated phonon intensity set by the Bose factor. The quasielastic width in all fits was fixed to a half width at half maximum (HWHM) of $\Gamma=3.00$~meV~\cite{LynnPRB2007}. Fit values are shown in Fig.~\ref{LCMO_fits}. The fits for the $\xi=0.25$ data at higher temperatures did not reliably converge due to the quasielastic component overwhelming the phonon peak and are excluded from this figure. Figure~\ref{LCMO_fits}(a) shows the phonon intrinsic width (HWHM) after deconvolution with the instrumental resolution displayed as a function of temperature. Fits from the data taken at zero field in the magnet are also included here, for $\xi=$0.15 and 0.20 at $T=$140~K and 265~K. The phonon peaks damp significantly for $\xi\geq0.15$, with the width increasing from roughly 0.5~meV at low temperatures to roughly 2.5~meV (for $\xi=0.20$) at temperatures exceeding $T_c$. When comparing to the damping reported \cite{MaschekPRB2016,MaschekPRB2018} in LSMO, we find that the damping near $T_c$ is roughly comparable in the two materials up to $\xi=0.15$ (as was also demonstrated in Fig.~\ref{Comparison}) but that by $\xi=0.20$ LCMO displays a high temperature damping ($\Gamma\approx2.5$~meV) in excess of any damping observed in LSMO (with a maximum reported value of $\Gamma\approx1.5$~meV). The change in phonon peak position (from $T=100$~K) is displayed in Fig.~\ref{LCMO_fits}(b) and shows that all phonon peaks soften with increasing temperature, with a typical softening of about 0.6~meV at 300~K. Previous work on La$_{0.75}$(Ca$_{0.45}$Sr$_{0.55}$)$_{0.25}$MnO$_3$ found transverse acoustic phonons near the polaron position that displayed a similar damping and softened by about 1~meV over this temperature range \cite{KiryukhinPRB2004}.

For the data at $\xi=0.25$ the temperature dependence was also measured for $\hbar\omega=2$~meV with smaller temperature increments. This is useful as a measure of the quasielastic intensity, as there will not be much contribution from the phonon at this energy. This temperature dependence, after subtracting the background, is shown in Fig.~\ref{TempDep}(a). The quasielastic intensity begins to rise when the temperature rises above 150~K, but the sharpest rise is observed just below $T_c$. The intensity increases modestly upon warming from $T_c$ to 315~K. Previously, the quasielastic scattering from dynamic polarons in LCMO had been found to persist to at least 400~K \cite{LynnPRB2007}. The equivalent data for La$_{0.70}$Sr$_{0.30}$MnO$_3$, measured at $\hbar\omega=3$~meV, is shown in Fig.~\ref{TempDep}(b). This rise is concentrated over a narrower temperature range, rising sharply above 275~K.

\begin{figure}
\centering
\includegraphics[width=8.3cm]{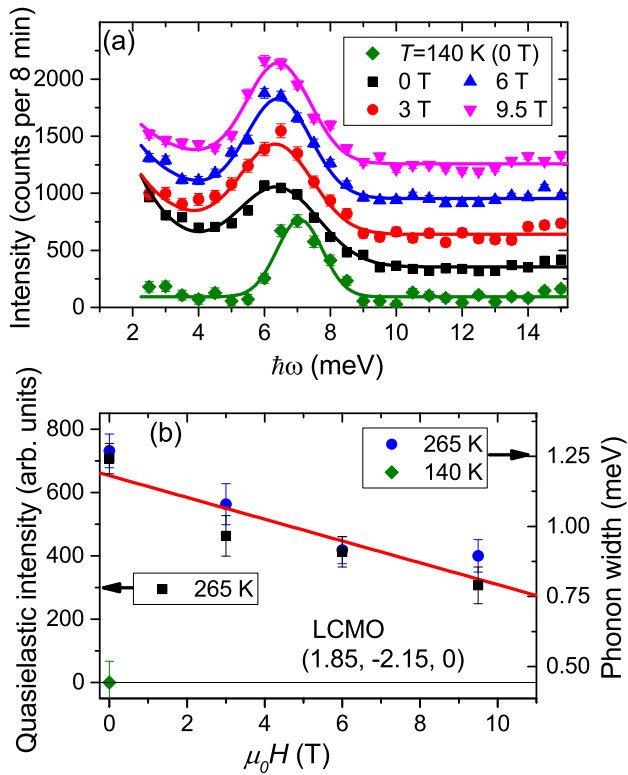} \vspace{-5mm}
\caption{(a) Constant-$Q$ scans in La$_{0.7}$Ca$_{0.3}$MnO$_3$ at $\vec{Q}=(1.85, -2.15, 0)$. Zero-field data were measured at 140~K and 265~K. Data in an applied field were measured at $T$=265~K. Successive data sets are offset for clarity. The lines are fits as described in the text. (b) Fit parameters for the quasielastic intensity and phonon width (HWHM).} \vspace{-2mm}
\label{Q0p15_Field}
\end{figure}

\begin{figure}
\centering
\includegraphics[width=8.3cm]{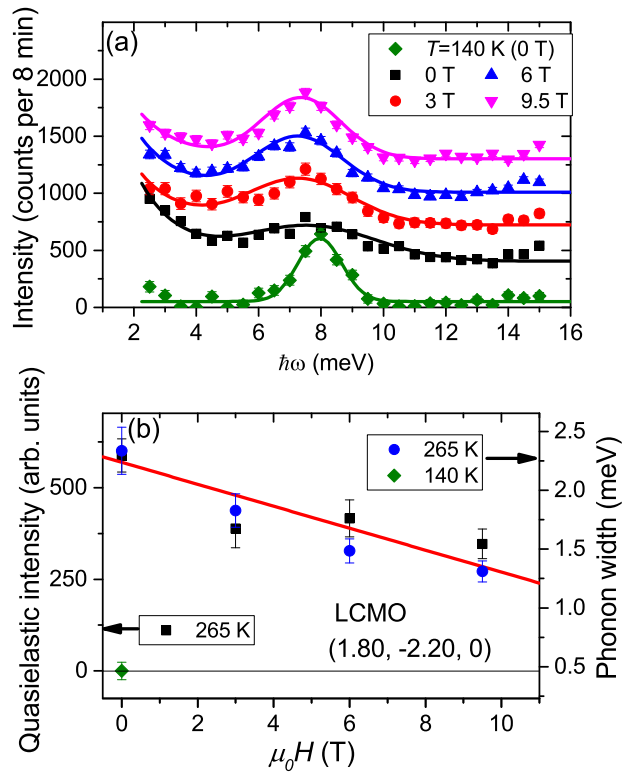} \vspace{-5mm}
\caption{(a) Constant-$Q$ scans in La$_{0.7}$Ca$_{0.3}$MnO$_3$ at $\vec{Q}=(1.80, -2.20, 0)$. Zero-field data were measured at 140~K and 265~K. Data in an applied field were measured at $T$=265~K. Successive data sets are offset for clarity. The lines are fits as described in the text. (b) Fit parameters for the quasielastic intensity and phonon width (HWHM).} \vspace{-2mm}
\label{Q0p20_Field}
\end{figure}

Figures~\ref{Q0p15_Field} and \ref{Q0p20_Field} show the data for LCMO at $\xi=0.15$ and $0.20$, respectively, measured in a superconducting magnet at $T$=140~K (zero field) and $T=265$~K (0~T, 3~T, 6~T, and 9.5~T). In each figure the data and fits are shown in the top panel while the fit values are shown in the lower panel. The fit values for the quasielastic intensity and phonon width are shown together with the axes chosen so that the 140-K fit values (for which the quasielastic intensity is fixed to zero) occur at the same vertical position. At 265~K, the application of a 9.5-T magnetic field decreases both the quasielastic intensity and the phonon damping by half compared to zero field. 

\begin{figure}
\centering
\includegraphics[width=8.3cm]{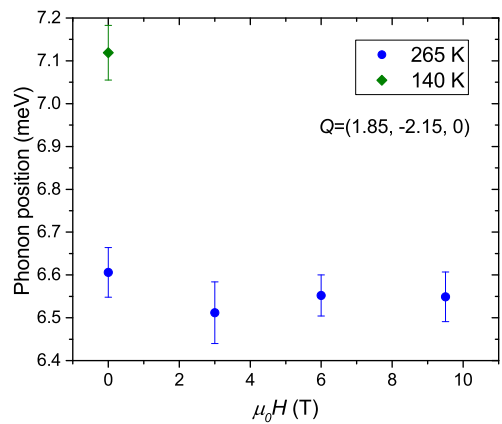} \vspace{-5mm}
\caption{Phonon position fit value for La$_{0.7}$Ca$_{0.3}$MnO$_3$ at $\vec{Q}=(1.85, -2.15, 0)$. Application of fields up to 9.5~T have no measurable effect on the phonon position at 265~K.} \vspace{-2mm}
\label{Field_position}
\end{figure}

In contrast, the field dependence of the phonon position fit value at $\vec{Q}=(1.85, -2.15, 0)$ is shown in Fig.~\ref{Field_position}. The phonon position softens by about 0.5~meV upon warming from 140~K to 265~K at zero field. However, applied fields up to 9.5~T do not reverse the softening, but rather have no measurable effect on the position.

\begin{figure}
\centering
\includegraphics[width=8.3cm]{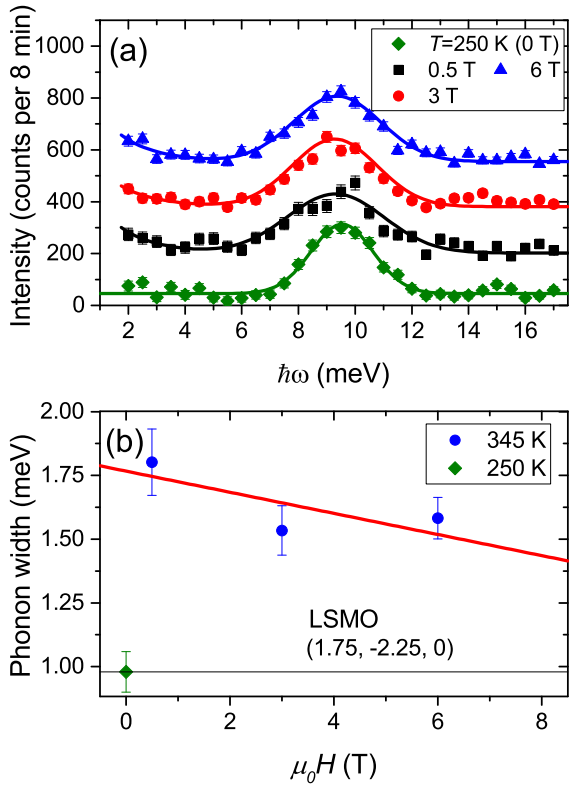} \vspace{-5mm}
\caption{(a) Constant-$Q$ scans in La$_{0.7}$Sr$_{0.3}$MnO$_3$ at $\vec{Q}=(1.75, -2.25, 0)$. Zero-field data were measured at 250~K. Data in an applied field were measured at $T$=345~K. Successive data sets are offset for clarity. The lines are fits as described in the text. (b) Fit parameters for the phonon width (HWHM).} \vspace{-2mm}
\label{LSMOfield}
\end{figure}

Figure~\ref{LSMOfield} shows the data for LSMO at $\xi=0.25$ measured at $T$=250~K (zero field) and $T=345$~K (0.5~T, 3~T, and 6~T). Near $T_c$ the low-field damping yields a phonon HWHM of about 1.75~meV which, while significant, is less than that observed in LCMO even at $\xi$=0.20. Conversely, this phonon is fairly broad (1 meV) even in the low-temperature data likely because of the elevated temperature in comparison to 140~K used as the low-temperature value in LCMO.  The applied magnetic field partially reverses the damping much as was observed in LCMO. The lines in Figs.~\ref{Q0p15_Field}(b), \ref{Q0p20_Field}(b), and \ref{LSMOfield}(b) have all been chosen to represent a field-dependent damping such that the width at 9.5~T is at the midpoint between the zero field width at the low and high temperatures.

\section{III. Discussion}

\begin{figure}
\centering
\includegraphics[width=8.3cm]{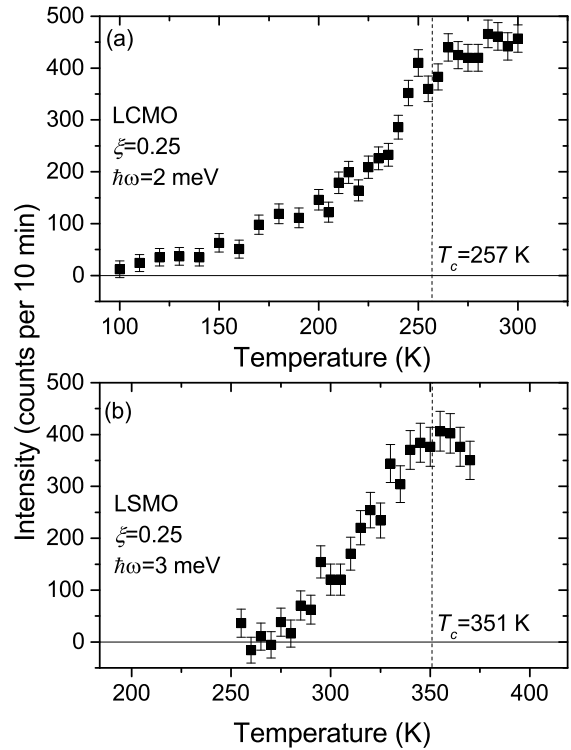} \vspace{-5mm}
\caption{(a) Quasielastic scattering intensity at $\hbar\omega=2$~meV and $\xi=0.25$ for La$_{0.70}$Ca$_{0.30}$MnO$_3$. (b) Quasielastic scattering intensity at $\hbar\omega=3$~meV and $\xi=0.25$ for La$_{0.70}$Sr$_{0.30}$MnO$_3$. These data consist primarily of scattering from the quasielastic component, as the phonon contribution will be quite small at these energies.} \vspace{-2mm}
\label{TempDep}
\end{figure}

Across a wide range of $R_{1-x}A_x$MnO$_3$ materials a trend is observed where compounds with a high $T_c$ typically display a weaker CMR effect. Therefore La$_{0.7}$Sr$_{0.3}$MnO$_3$ ($T_c=$351~K) \cite{UrushibaPRB1995} displays a significantly less pronounced CMR effect than La$_{0.7}$Ca$_{0.3}$MnO$_3$ ($T_c=257$~K) \cite{SchifferPRL1995}. This was historically taken as evidence that LSMO featured weaker electron-lattice coupling and therefore displayed weaker CE-type lattice distortions in the high-temperature phase. However, more recent work has confirmed that LSMO displays lattice distortions at the transition temperature comparable to those observed in LCMO and other compounds with a significant CMR effect \cite{WeberPRB2013} as well as evidence of dynamic CE-type polarons \cite{MaschekPRB2016}. Despite the comparable lattice distortions, we find that the quasielastic scattering arising from short-lifetime dynamic polarons near $T_c$ is a factor of 5 more intense in LCMO than in LSMO. This quasielastic intensity rises sharply as the temperature approaches $T_c$ and remains significant at temperatures above $T_c$ up to at least 400~K \cite{LynnPRB2007}. This is comparable to the temperature dependence of the diffuse scattering arising from uncorrelated static polarons in this material \cite{AdamsPRL2000}. We compare the temperature dependence of the quasielastic scattering in LCMO with that in LSMO, and find that the rise in quasielastic intensity occurs over a larger temperature range in LCMO. Figure~\ref{LSMO_0p20_0p25} shows that the quasielastic intensity in LSMO is quite low up to 300~K ($\approx0.85T_c$) and rises most sharply between 300~K and 325~K. The quasielastic scattering in LCMO, as observed in Fig.~\ref{LCMO_Data}, rises steadily starting near 150~K ($\approx0.58T_c$). Figure~\ref{TempDep} shows these temperature dependencies with equally wide scales and the transition temperatures aligned. Recently, La$_{0.8}$Sr$_{0.2}$MnO$_3$ was also shown to feature a quasielastic component that appeared over a narrow temperature range \cite{MaschekPRB2018}.

The transverse acoustic phonons in LCMO are found to damp with rising temperature for $\xi=$0.15 and 0.20, approaching the polaron position. Similar to the quasielastic intensity, the damping increases upon warming through $T_c$ but remains fairly constant as the temperature is increased further. LCMO undergoes significant local structural changes at $T_c$, with changes in the lattice parameters \cite{RadaelliPRL1995}, increases in the O and Mn atomic displacements \cite{DaiPRB1996}, and the width of the distribution of Mn-O distances \cite{BridgesPRB2010,BoothPRL1998}. It is natural to associate this damping with increased scattering of phonons from the local distortions surrounding polarons. The temperature dependence of the phonon damping is particularly reminiscent of the temperature dependence of the variance in Mn-O bond distances ascribed to polarons \cite{BridgesPRB2010}. The transverse phonons in LSMO are also found to damp with increasing temperature in agreement with previous result s\cite{MaschekPRB2016}. The high temperature damping is fairly similar between the two materials for $\xi=0.15$. However, the damping becomes more pronounced in LCMO at larger reduced wave vectors so that the damping near $T_c$  is larger in LCMO at $\xi=0.20$ (2.5 meV) than in LSMO even at the larger reduced wave vector of $\xi=0.25$ (1.75~meV).

The application of a magnetic field lowers both the phonon damping and the quasielastic scattering; however, a field of 9.5~T is sufficient to return these parameters only half way to their low-temperature values. This is in keeping with previous information about the effect of a magnetic field on the number of polarons present. Ridges of magnetic scattering arising from polaron-mediated spin correlations follows a similar field dependence \cite{HeltonPRB2014}. Near $T_c$ an applied 9-T field lowered the variance in Mn-O bond lengths about half way back to its low-temperature value \cite{BridgesPRB2010}. Correlations between static polarons are suppressed by a much smaller magnetic field of about 5~T \cite{LynnJAP2001}, comparable to the fields under which CMR is observed \cite{SchifferPRL1995}, and a field of only 1~T will suppress these correlations in a bilayer manganite~\cite{VasiliuDolocPRL1999}. These facts have been used to argue that colossal magnetoresistance follows polaron correlations rather than the number of polarons.

Transverse acoustic phonons in LCMO at all measured wave vectors soften upon warming, with phonon peaks decreasing in energy by about 0.6~meV from 100~K to 300~K. We observe comparable softening in LSMO from 250~K to 370~K, although a previous work had found significantly more softening in that material at wave vectors close to the polaron position \cite{MaschekPRB2016}. The softening is continuous, with no apparent anomaly at $T_c$. A slightly larger softening observed in La$_{0.75}$(Ca$_{0.45}$Sr$_{0.55}$)$_{0.25}$MnO$_3$ likewise featured no anomaly with temperature and could not be explained by lattice thermal expansion without a nonphysically large Gr\"{u}neisen constant \cite{KiryukhinPRB2004}. Most notably, the softening is field independent up to 9.5~T at 265~K. Colossal magnetoresistance in the manganites has often been analyzed as a percolation effect \cite{UeharaNature1999,MayrPRL2001,BaldiniPNAS2015} with a precipitous breakdown of conducting pathways occurring upon warming through $T_c$. The distribution width of Mn-O bonds has been found to display a universal relationship with magnetization for fields above 2~T with two distinct regimes; for sample magnetization below about 65\% of saturation the change in structural distortion with magnetization was found to be small \cite{BridgesPRB2010}. The magnetization of La$_{0.7}$Ca$_{0.3}$MnO$_3$ at 265~K and 9~T was reported in that study to be about 70\% of saturation, indicating that our data are mostly still in the low-field regime. One explanation for these results could be that phonon softening is driven by the relative fraction of insulating clusters in the sample while applied fields in this regime primarily re-orient clusters of conducting and insulating phases with a relatively smaller effect on their relative fractions.

\section{IV. Summary}

Transverse acoustic phonons in La$_{0.7}$Ca$_{0.3}$MnO$_3$ measured near the (2,~2,~0) Bragg position damp and soften with increasing temperature. The damping remains relatively constant above $T_c$ and becomes more pronounced at wavevectors approaching the polaron position, with a maximum measured width (HWHM) of 2.5~meV at $\xi=0.20$ above $T_c$. Applied magnetic fields up to 9.5~T only partially reverse the damping, lowering the phonon width at high temperature. The temperature and magnetic-field dependence of the damping are similar to those of the distribution width of Mn-O bond distances measured by x-ray pair distribution function \cite{BridgesPRB2010}, strongly suggesting that the phonon damping arises from scattering from local Jahn-Teller distortions surrounding polarons.

In addition to the phonons, a quasielastic component arising from local distortions is observed and fit with a width of $\Gamma=3.00$~meV \cite{LynnPRB2007}. When compared to La$_{0.7}$Sr$_{0.3}$MnO$_3$, a similar compound with a higher transition temperature and significantly smaller CMR effect but comparable lattice distortions, the quasielastic component is about five time stronger in LCMO and features a temperature dependence that rises over a less concentrated temperature range. The magnetic field dependence of the quasielastic component closely matches that of the phonon damping. Phonon softening is observed at all wave vectors. While no anomaly is observed in the softening, the softening is likely too strong to be explained solely by thermal expansion. An applied magnetic field does not affect the phonon energy, possibly suggesting an effect that arises from relative concentrations of coexisting metallic and insulating phases and is insensitive to details of percolation pathways.

\section{Acknowledgement}

J.S.H acknowledges partial support from the Office of Naval Research through the Naval Academy Research Council.

emails: $\ast$ helton@usna.edu, $\dag$ jeffrey.lynn@nist.gov\\


\bibliography{LCMOPhonons}

\end{document}